# Design of the HARMONI Pyramid WFS module

Noah Schwartz[a*], Jean-François Sauvage[b], Edgard Renault[d], Carlos Correia[c], Benoit Neichel[d], Thierry Fusco[b], Kjetil Dohlen[d], Kacem El Hadi[d], Cyril Petit[b], Elodie Choquet[d], Vincent Chambouleyron[d], Jerome Paufique[e], Fraser Clarke[f], Niranjan Thatte[f], Ian Bryson[a]

[a]UK Astronomy Technology Centre, Blackford Hill, Edinburgh EH9 3HJ, United Kingdom; [b]ONERA, 29 avenue de la Division Leclerc, 92322 Châtillon, France; [c]W. M. Keck Observatory, 65-1120 Mamalahoa Hwy, Kamuela, Hawaii 96743, USA; [d]Marseille Univ, CNRS, LAM, Laboratoire d'Astrophysique de Marseille, Marseille, France; [e]European Southern Observatory, Karl-Schwarzschild-Str. 2, 85748 Garching, Germany; [f]Dept. of Astrophysics, University of Oxford, Keble Road, Oxford, OX1 3RH, United Kingdom;

## ABSTRACT

Current designs for all three extremely large telescopes show the overwhelming adoption of the pyramid wavefront sensor (P-WFS) as the WFS of choice for adaptive optics (AO) systems sensing on natural guide stars (NGS) or extended objects. The key advantages of the P-WFS over the Shack-Hartmann are known and are mainly provided by the improved sensitivity (fainter NGS) and reduced sensitivity to spatial aliasing. However, robustness and tolerances of the P-WFS for the ELTs are not currently well understood.

In this paper, we present simulation results for the single-conjugate AO mode of HARMONI, a visible and near-infrared integral field spectrograph for the European Extremely Large Telescope. We first explore the wavefront sensing issues related to the telescope itself; namely the island effect (i.e. differential piston) and M1 segments phasing errors. We present mitigation strategies to the island effect and their performance. We then focus on some performance optimisation aspects of the AO design to explore the impact of the RTC latency and the optical gain issues, which will in particular affect the high-contrast mode of HARMONI.

Finally, we investigate the influence of the quality of glass pyramid prism itself, and of optical aberrations on the final AO performance. By relaxing the tolerances on the fabrication of the prism, we are able to reduce hardware costs and simplify integration. We show the importance of calibration (i.e. updating the control matrix) to capture any displacement of the telescope pupil and rotation of the support structure for M4. We also show the importance of the number of pixels used for wavefront sensing to relax tolerances of the pyramid prism. Finally, we present a detailed optical design of the pyramid prism, central element of the P-WFS.

**Keywords:** ELT, wavefront sensing, SCAO, pyramid, modelling, optical design

## 1. INTRODUCTION

The scientific advances to be expected from the ability to operate a 40-m-class telescope at its diffraction limit form the core of the Science Case for the ELT. The need for diffraction-limited performance is hence a strong driver for the design of the telescope and its associated instruments. Among the science cases defined for HARMONI, some require the use of a high-performance SCAO system, for example: solar system, high contrast spectroscopy of planetary mass companions to nearby stars, intermediate mass black holes.

HARMONI[1,2] is a visible and near-infrared integral field spectrograph, providing the ELT's core spectroscopic capability. It will exploit the ELT's scientific niche in its early years, starting at first light. To get the full sensitivity and spatial resolution gain, HARMONI will work at diffraction-limited scales. This will be possible using two complementary adaptive optics systems.

*noah.schwartz@stfc.ac.uk; phone +44 131 668 8256

The first is a simple yet high-performing Single Conjugate AO (SCAO) system (i.e. excellent performance, low sky coverage); the second is a Laser Tomographic AO (LTAO) system (i.e. medium performance, very good sky coverage). In section 2 we present a brief overview of HARMONI and its SCAO module. In section 3, we first investigate the contribution of the telescope on SCAO performance and on the specific point of the mitigation of the island effect. In section 4, we then concentrate on SCAO system design itself to optimise performance by investigating the impact of RTC latency on SCAO performance and the strategy for correcting optical gain issues. Finally, in section 5 we present HARMONI's strategy to help reduce the impact of the glass prism defects on performance and the optical design.

## 2. HARMONI

### 2.1 Natural Guide Star System

Figure 1 shows a schematic of the HARMONI instrument where the Natural Guide Star System (NGSS) is the host of the SCAO module, which is described in the following section. The LGS-WFS module contains 6 LGS wavefront sensors providing HARMONI with its LTAO capability. The focal plane relay module is a modified Offner relay providing a 2-arcminute field to the cryostat and to the NGS sensing module. Finally, the NGSS module provides the SCAO, high-contrast, LTAO-NGS and seeing-limited guiding capabilities to HARMONI.

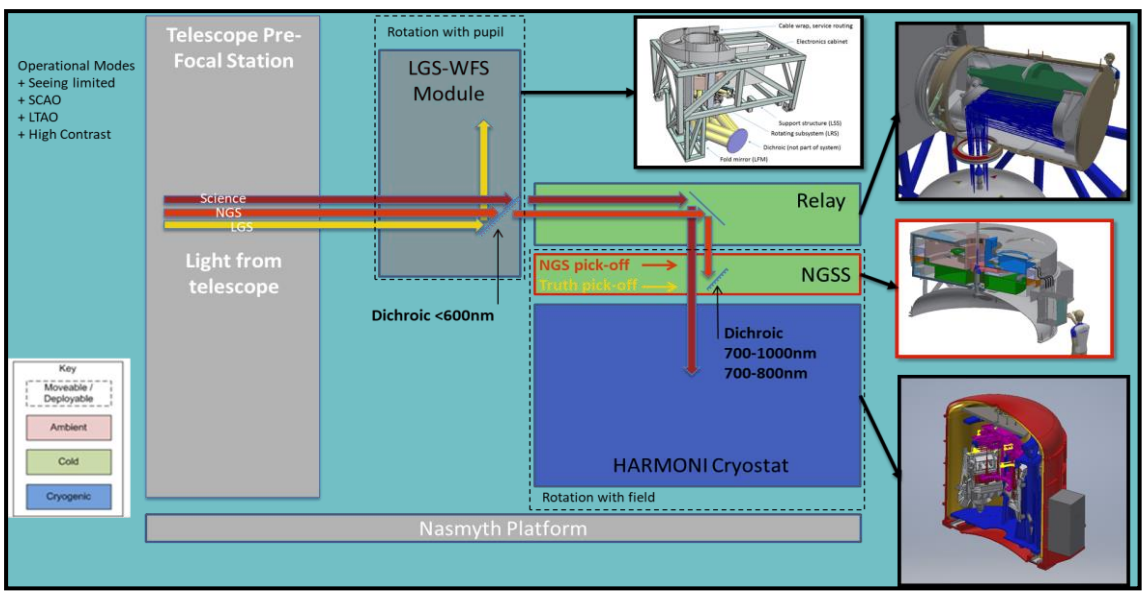

Figure 1: Schematic of the HARMONI instrument showing the different modules.

### 2.2 SCAO Module

Figure 2 shows a CAD model of the NGSS module. It is mounted directly onto the cryostat of the Integral Field Spectrograph (IFS) to maximize stability to the science focal plane, and is cooled at -20C to match the focal plane relay system. First, the 2-arcminute field provided by the relay is split by the NGSS dichroic between the science and the sensing wavelengths. To optimize both science and AO performance, the NGSS dichroics have two different spectral cut-offs at 700-800 and 800-1000nm respectively (TBC). The object selection mirror (OSM) allows the SCAO module to use a NGS up to 15 arcsec away from the science target. A Low Order Loop[3] – composed of a low-order deformable mirror (LODM) and a low-order Shack-Hartmann (BlueSH) - allows for the NCPA compensation and stabilisation. A linear Atmosphere Dispersion Corrector (ADC) minimised the chromatic errors. Finally, a WFS based on a 100x100 pyramid is coupled to a fast and low noise detector (an OCAM2 camera running from 500 Hz to 1 kHz).

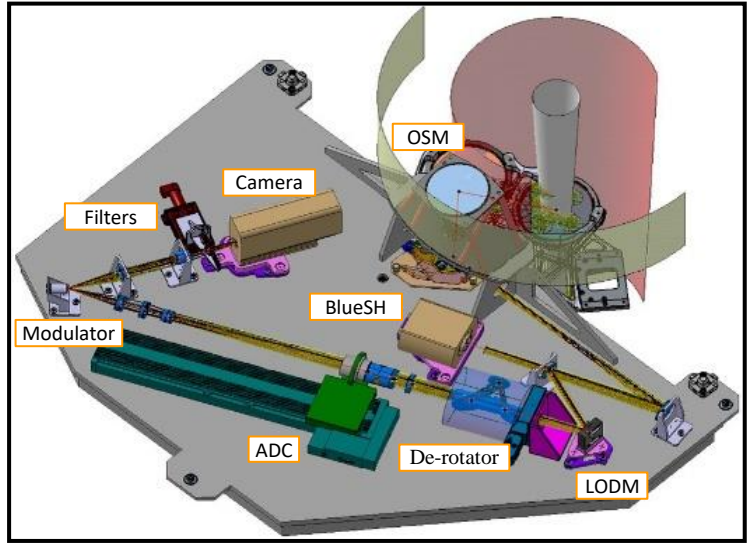

Figure 2: Schematic of the NGSS module.

As per defined in the requirement specifications, the SCAO mode shall deliver at least 70% SR in K-band under median seeing conditions, and a reasonable performance over all seeing conditions and up to 1.2 arcsec. Figure 3 shows the overall performance of the SCAO module with respect to NGS magnitude. The global performance estimation only accounts for the SCAO pure turbulence residual, the telescope error budget, and the island effect residuals. The dynamic contributions of the telescope are mainly driven by wind shake (1.4'' residual for to the worst case), static telescope contribution (30 nm RMS), and islands as detailed in section 4, 50 nm for high turbulence and 35 nm for nominal turbulence.

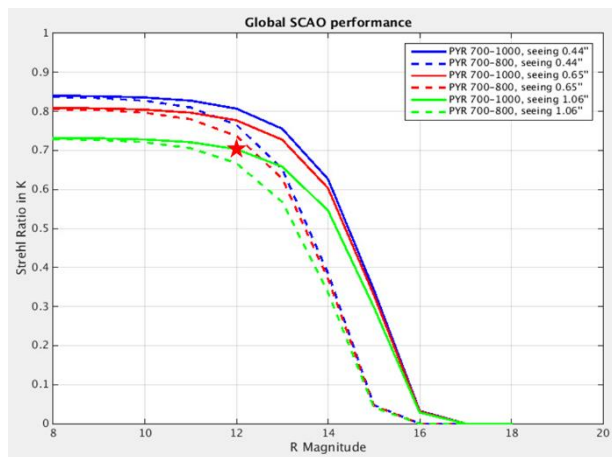

Figure 3: Global SCAO performance, accounting for SCAO pure turbulence residual, telescope error budget, and island effect residual. Only P-WFS cases are considered here (i.e. no SH-WFS comparison) under median turbulence conditions.

## 3. TELESCOPE CONTRIBUTION

All results in this paper are obtained using Object-Oriented Matlab Adaptive Optics Simulator, OOMAO[4]. OOMAO is an AO numerical environment embedding tailored features for: a) Monte-Carlo simulations, b) analytic studies, and c) driving optical experiments interfacing seamlessly with bench setups. It is one of the most commonly used numerical models. Its object-oriented structure gives it a great flexibility whilst benefiting from Matlab built-in hardware-acceleration support (e.g. parallel computing on multi-cores, GPU acceleration, and vectorisation).

## 3.1 Impact of M1 residual errors

In this section, we present results of end-to-end simulations that include static high spatial frequency residual errors that are expected from the primary mirror M1 of the ELT. These errors can potentially be extremely problematic for SCAO as they present phase discontinuities that cannot easily be corrected by the SCAO system. M1 phasing errors contains 10 different terms: M1 polishing errors, main structure gravity, main structure temperature, phasing errors, scalloping errors, segment AIV, segment coating stress, segment support, and segment temperature effects (long and short term).

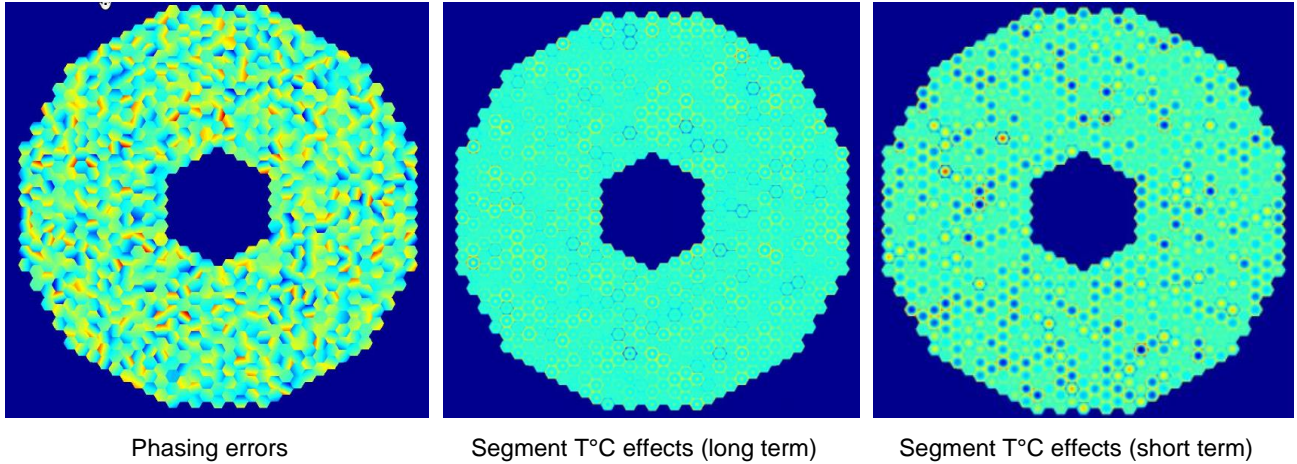

Phasing errors      Segment T°C effects (long term)      Segment T°C effects (short term)

Figure 4: Phase map representing the impact of segment phasing (i.e. piston-tip-tilt errors) and segment temperature errors.

We focus on the three most important terms exhibiting the largest phase error (see Figure 4), namely phasing error and segment temperature effects. Simulations were performed in median seeing (0.65''), without noise and for at least 1000 iterations. The final performance was estimated by the residual RMS error over the full pupil and the residual phase map and stability of AO loop were also checked.

Table 1: Summary of the impact of static M1 aberrations on SCAO performance using the anticipate error amplitudes.

| Type of error | M1 error [nm RMS] | Res. Error [nm RMS] | Quad. Diff. [nm RMS] | SR(H) [%] | SR(K) [%] |
|---|---|---|---|---|---|
| **No static errors** | 0 | 103 | - | 85.8 | 91.6 |
| **Phasing** | 36 | 105 | +23 | 85.2 | 91.2 |
| **T°C (long term)** | 21 | 106 | +26 | 85.0 | 91.1 |
| **T°C (short term)** | 24 | 104 | +16 | 85.5 | 91.4 |
| **All three combined** | 48 | 109 | +37 | 84.2 | 90.5 |

No significant impact of M1 aberrations is noticeable. Some errors terms (e.g. phasing errors) are somewhat corrected by the AO loop – a small reduction in the amplitude of the error is seen. After a certain phasing errors level is reached (much larger than the estimated errors provided by ESO) the SCAO loop actually starts amplifying the error, but the impact is still very small. Other errors (e.g. temperature effects) are not well corrected, but are not really amplified by the AO loop either. The input error is approximately the same as the output error after AO correction. For all cases, there is no significant amplification of the M1 errors by the AO loop. A combination of multiple errors also shows very similar trends to errors taken individually. This study shows that - in median turbulence conditions JQMedian - static residual phasing errors of M1 do not affect significantly SCAO performance.

## 3.2 Impact of telescope spider

One of the major challenges to meet the overall performance requirements is the presence of telescope spiders. They namely create two detrimental effects: the island effect and the low-wind effect[5]. The island effect[6, 7, 8] is a well-known phenomenon appearing in AO system with spiders wider than Fried parameter and wider than the equivalent sub-aperture size.

As the telescope design fixes the spiders' location and size there is very little we can change about their thicknesses or any other of their characteristics. On the other hand, M4 actuators are driven into position to a known and absolute distance from the common reference body: actuator extensions across the entire DM surface shall be available. It is therefore possible to modify the behaviour of the DM by acting on the extension of the actuators. In particular, we can emulate the behaviour of a DM with a continuous surface by coupling edge actuators. The coupling is done for all pairs of actuators directly opposite of each other from either side of the spiders and on two different DM petals.

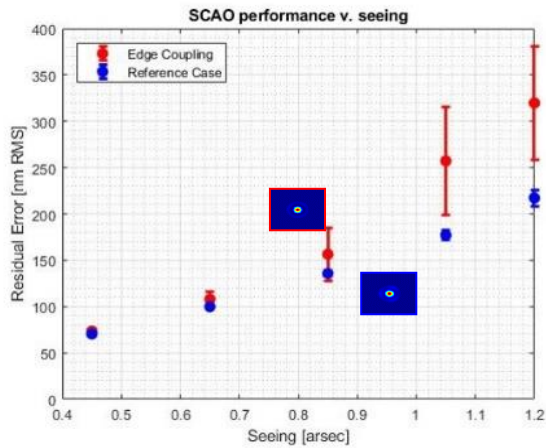
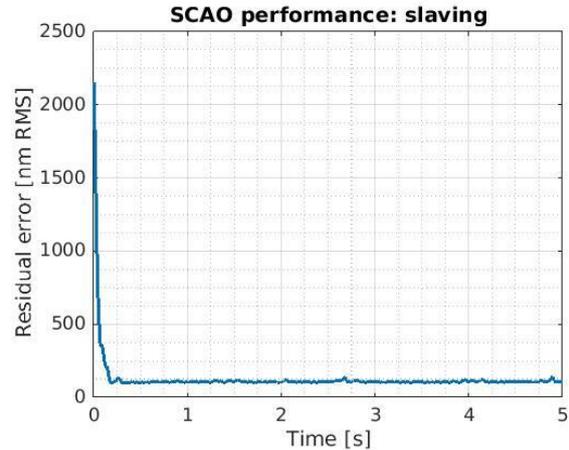

Figure 5: Impact of the edge actuator coupling on performance as a function of seeing.

Figure 6: SCAO residual error as a function of time for 0.65" at 30° zenith angle.

Figure 5 SCAO performance as a function of seeing condition. It compares the nominal residuals (i.e. circular aperture, with only atmospheric contributions) to the island effect correction using the Edge Actuator Coupling (EAC) technique residuals (i.e. ELT segmented aperture, with only atmospheric turbulence). This is to be compared to thousands of nm of residual error obtained without EAC. This shows a great reduction in the final residual error and a reasonable increase of this error as the seeing degrades. This makes this technique compatible with high-contrast imaging - when seeing conditions are good – as only a very small additional error is present. Figure 6 show the residual error as a function of time demonstrating the stability of the technique.

## 4. SCAO MODULE – PERFORMANCE OPTIMISATION

### 4.1 Impact of RTC latency

Initially the loop framerate was specified to 500 Hz, even if a higher performance was specified as a goal. In this section, we detail the temporal error budget and show that a 1 kHz framerate with a latency as close as possible to 2 frames for the control loop helps in two ways: 1/Improve the performance for the high-contrast case; 2/ Give a risk mitigation (margin) to cope with the unknown telescope environment (e.g. unexpected vibrations).

Benchmarking of the RTC proposed by the University of Durham[9,10] has showed that a 1 kHz framerate is achievable with a single core computer, but the 2 frames delay (2 ms) is still under investigation. A double core solution might be required. We define the RTC latency measured from the moment when the RTC receives the last pixel from the WFS camera to when the DM command is ready to be sent.

**Bandwidth analysis**

We compute Table 2 the AO correction bandwidth corresponding to different configurations. We compare different framerates (namely 500 Hz and 1 kHz), a 1 ms RTC latency as measured in the current RTC benchmarking, and two reduced RTC latencies 0.3 ms and 0.1 ms. The WFS readout time is 0.484 ms – a number provided by camera manufacturer FirstLight - for the actual measured performance of the OCAM2 camera. ESO provided the number for the Central Control Software (CCS) and communication delay (comm. delay).

The main result is that running the AO loop at 1 kHz with 3 frames delay does not help in increasing the overall bandwidth of the system; compared to 500 Hz with 2 frames delay. The bandwidth is only increased by 5 Hz, from 32 Hz to 37 Hz. An improved global latency of at least 2.5 frames is required to improve the performance and provide a >50 Hz bandwidth, and reaching a 2.3 frames delay allow to reach 57 Hz.

Table 2: Bandwidth computed for different framerates and RTC latencies. All numbers are expressed in ms except where indicated. This table assumes a phase margin of 45°.

|     |                            | 500   | 1000  | 1000  | 1000  |
|-----|----------------------------|-------|-------|-------|-------|
|     | **Frame rate (Hz)**        |       |       |       |       |
|     | **Temporal contributors**  |       |       |       |       |
| **WFS** | **WFS integration (ms)**   | 2     | 1     | 1     | 1     |
|     | **WFS read out (ms)**      | 0,484 | 0,484 | 0,484 | 0,484 |
|     | **Comm. delay (ms)**       | 0,03  | 0,03  | 0,03  | 0,03  |
| **RTC** | **RTC latency (ms)**       | 1     | 1     | 0,3   | 0,1   |
|     | **Comm. delay (ms)**       | 0,03  | 0,03  | 0,03  | 0,03  |
| **CCS** | **E-ELT CCS computation (ms)** | 0 | 0 | 0 | 0 |
|     | **Comm. delay (ms)**       | 0,03  | 0,03  | 0,03  | 0,03  |
|     | **M4LCS computation (ms)** | 0,25  | 0,25  | 0,25  | 0,25  |
|     | **M4 command application (ms)** | 0,4 | 0,4 | 0,4 | 0,4 |
|     | **total (ms)**             | 4,224 | 3,224 | 2,524 | 2,324 |
|     | **Global delay (frame)**   | 2,112 | 3,224 | 2,524 | 2,324 |
|     | **Bandwidth (Hz)**         | **32,0** | **37,4** | **51,0** | **57,1** |

**E2E simulations**

A set of end-to-end simulations were launched for a nominal configuration: a median turbulence condition, modulation 3 $\lambda/D$, 10.5 guide star magnitude. Two different wind profiles are considered with altitudes of phase screens of h = [609, 10409, 19426] m, and a Fried parameter $r_0$ = 14.26 cm: 1/ $V_{wind}$ = [5.5, 5.5, 5.1] m/s, a coherence time of $\tau_0$ = 13.8 ms, and 2/ $V_{wind}$ = [11, 5.5, 5.1] m/s, a coherence time of $\tau_0$= 7.59 ms. The only varying parameters are the AO loop framerate and the loop delay. Integer numbers of delay frames are introduced with OOMAO in this simulation, from 2 to 4 frames delay are considered for each condition. The residual WFE is plotted for each condition in Figure 7.

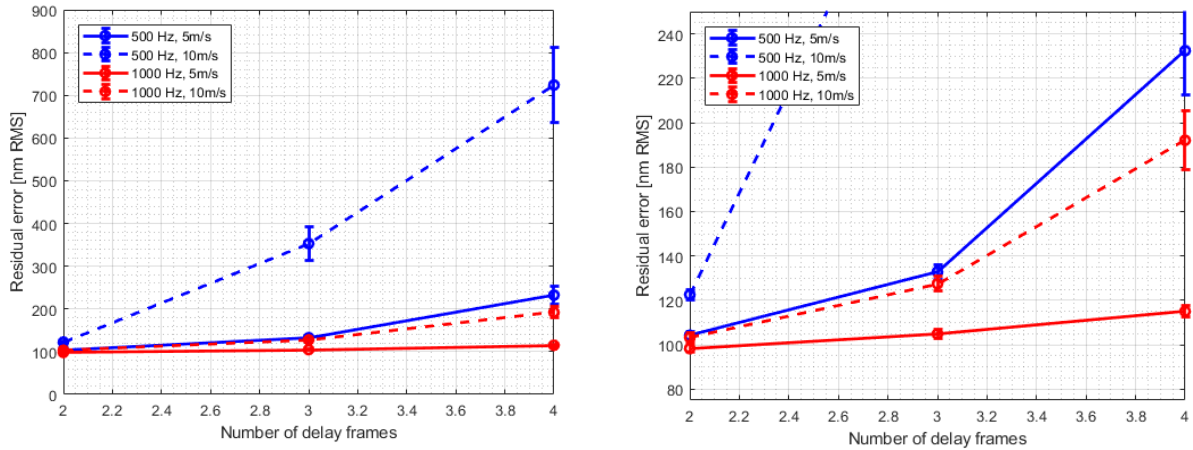

Figure 7: Performance (and standard deviation to show spread of results) as a function of delay in frames and loop sampling rate. Right: Zoom to show small residual errors from 80 to 240 nm RMS only.

For a framerate of 500 Hz and a wind speed of 10 m/s a latency of 2 frames is required to achieve the AO residual specification. A 3-frame delay would increase the error to 350 nm RMS residual WFE, which is not acceptable. For a framerate of 1 kHz, a delay of 3 frames translates into a residual of 126 nm RMS, which is slightly worse than the case of 500 Hz, 2 frames delay (122 nm RMS). In order to take the full benefit of running the AO loop at 1 kHz, it is mandatory to reduce the latency significantly below 3 frames. Under these conditions (i.e. 1 kHz & 2 frames), the residual wavefront error can be further decreases from 122 nm to 101 nm RMS ground wind speed as high as 10 m/s. Table 3 presents a summary of the main results for 500 Hz and 1 kHz.

Table 3: Summary of performance for 500 Hz and 1 kHz for 2 and 3 frames delay. No noise.

| Frame Rate | Frame delay | Res. Error [nm RMS] | Quad. Diff. [nm RMS] | SR(H) [%] | SR(K) [%] |
|---|---|---|---|---|---|
| **500 Hz** | 2 | 122 | - | 80.7 | 88.4 |
| **1 kHz** | 2 | 101 | -68 | 86.3 | 91.8 |
| **1 kHz** | 3 | 126 | +33 | 79.5 | 87.6 |

**Impact for High Contrast performance**

The benefit of reaching 1 kHz with 2 frames delay is shown on Figure 8 for the radially averaged residual phase power spectral density (PSD), and on Figure 9 PSD in 2D maps. The residual PSD shows two major points:

- Running the loop at 1 kHz with 3 frames delay is useless in terms of residual WFE. The PSD are almost superimposed, and even slightly worse in the 1 kHz case for very low orders.
- A clear improvement on the entire corrected area of the focal plane is visible (simulation without noise), with potentially a factor 5 improvement for high-contrast. In particular, the temporal error residual (aligned with the wind direction in Figure 9 left) is attenuated at a higher framerate, 2 frames delay (right).

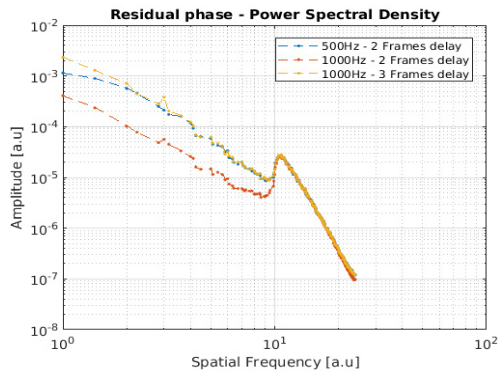

Figure 8: Residual power Spectral Density for 500 Hz & 1 kHz.

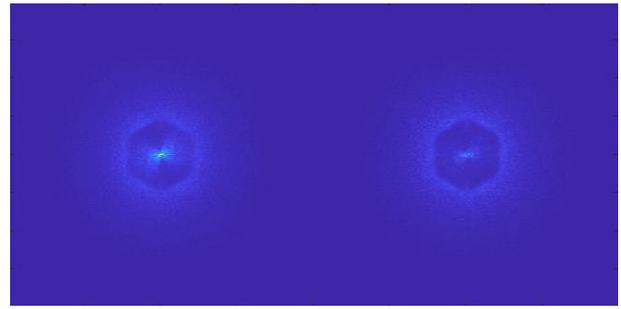

Figure 9: Power spectral density (PSD) with 2 frames delay. (Left) 500 Hz, (right) 1 kHz. Both simulations are with 2 frames delay, which is probably optimistic for the 1 kHz.

## 4.2 Optical gain correction strategy

**Definition of the optical gain**

The P-WFS suffers from a well-known drawback in typical operations: an overall modification of the sensor's response and a reduction of the measurement sensitivity. This is called the optical gain[11, 12] (OG), and is a major issue in terms of wavefront sensing performance and operability for AO systems operating on the ELT. When operating outside its calibration regime – i.e. with residual uncorrected residual turbulence ($SR_{RES} < 100\%$) - the P-WFS experiences a loss in sensitivity due to the non-linear nature of the output signal. The OG gain issues can be mitigated by applying compensatory gains – generally one for each controlled mode - known as modal gain compensation.

$$G_{opt} = \frac{\langle M^{sensing}(\varphi_{res}) | M^{calib}(\varphi_{res}) \rangle}{\| M^{calib}(\varphi_{res}) \|^2} \quad (1)$$

The definition of the optical gain $G_{opt}$ is given equation (1), with M the interaction matrices respectively for the calibration and a given sensing working points (i.e. with residual phase errors), and $\varphi_{res}$ the residual phase. Figure 10 gives a generic illustration of the non-linear P-WFS response[13]. The actual measurement depends on the modulation radius, the considered mode (e.g. Zernikes or Karhunen-Loève), and on the amount of residual aberrations (i.e. working point). Figure 10 also gives an illustration of the difference in PSF shape has seen at the apex of the pyramid prism during calibration and during operations due to the presence of residual aberrations.

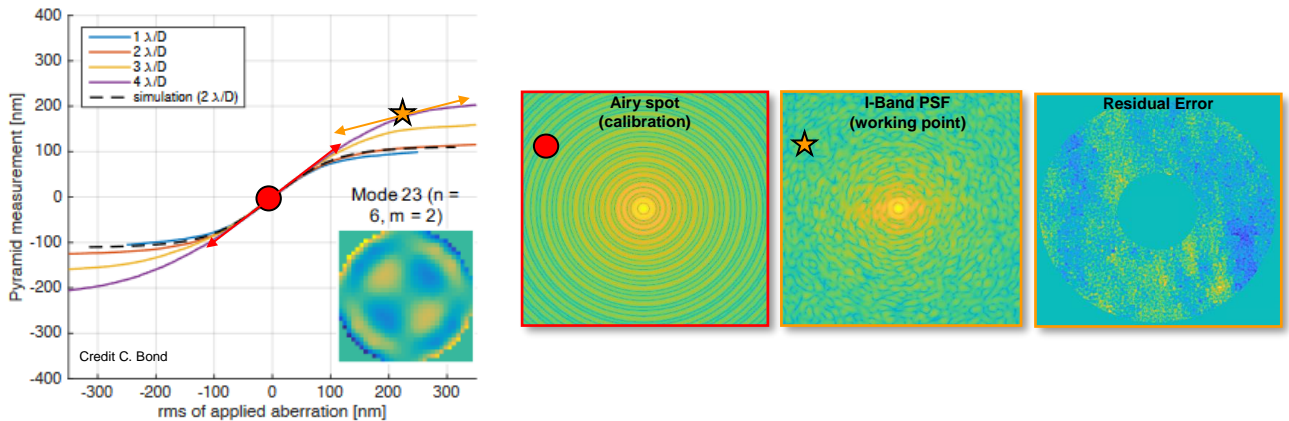

Figure 10: Illustration of the non-linear P-WFS response and of the loss of sensitivity associated with a working point (yellow star) outside the calibration regime (red dot) containing residual turbulent wavefront aberrations.

**Implementation strategy**

Three difference options are being simultaneously investigated for HARMONI: 1) Scalar gain[6], 2) dithering[14] and 3) analytical model[12]. During PDR we investigated the use of a unique scalar gain applied to all modes simultaneously and scaled only by the seeing condition. This strategy has the main advantages of meeting the SCAO requirements while being extremely straightforward to implement. OG has in practice only a limited impact on residual error for HARMONI's SCAO overall performance[3], [6]. However, a good control of the OG is essential for High-Contrast imaging[15]. OG is also changed when the P-WFS is operating at non-zero offsets, such as for Non-Common Path Aberrations (NCPA) compensation. In HARMONI, the use of the Low-Order Loop will mitigate the impact of low-order NCPAs on final SCAO performance[3] and will require in addition the P-WFS to operate with non-zero offsets.

For this reason, two other options are being investigated in parallel. Dithering - the PDR strategy - relies on an additional tip-tilt modulation of the P-WFS modulator to extract the optical gain of tip-tilt. An analytical scaling law (e.g. based on look-up tables) is then applied to obtain the final modal gain compensation at each iteration step. This has the main advantages of being able of tracking the ever-changing turbulence, of adding minimal disruptions to the system, and of having been demonstrated in simulations and on sky for smaller AO systems. We are currently estimating its performance HARMONI's SCAO module. Finally, an analytical model based on a convolutive model of the P-WFS is being developed. It requires only minimal input information with no disruption to the system, can track the ever-changing turbulence in real-time, and is computationally fast.

Summary of the modal gain compensation strategies:

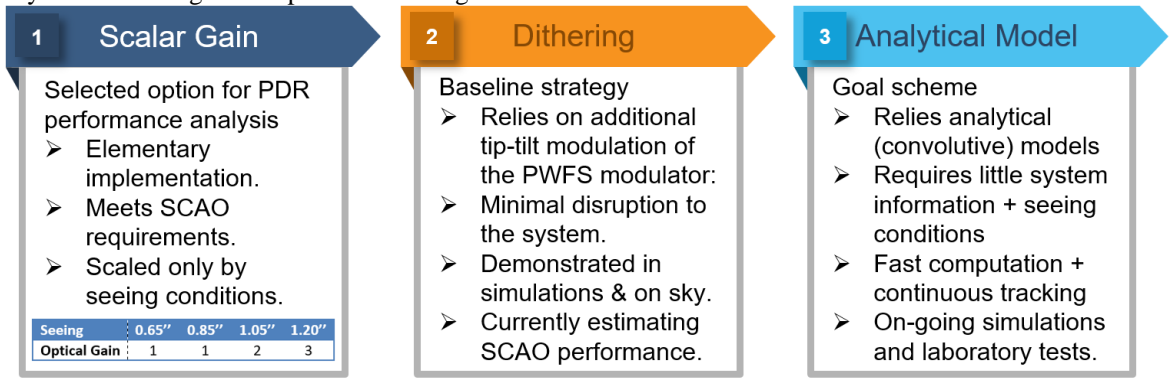

Figure 11 shows the modal compensation gain as a function of Karhunen-Loève for different seeing conditions. First KL modes show change in OG values due to the P-WFS's modulation radius. Modal compensation of the optical gains brings some marginal performance increase over global scalar compensation (work in progress) but have the main attraction of being able to deal with NCPA and improving High-contrast imaging.

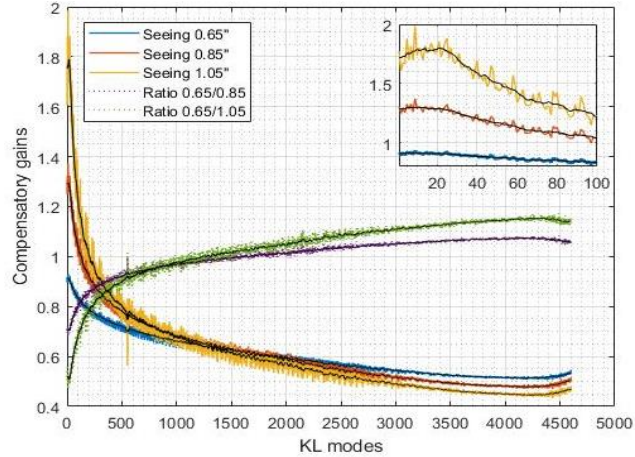

Figure 11: Modal compensation gain as a function of Karhunen-Loève for different seeing conditions. Inset shows a zoom for KL modes between 1 and 100. The solid black lines are a visual aid showing a rolling average of the modal gains. Green and purple represent the ratio between the modal compensation gains for 2 different seeing conditions.

## 5. PYRAMID OPTICAL DESIGN

### 5.1 Impact of pyramid faces orthogonality

The optical quality of the glass pyramid prism (e.g. roof, sharpness of vertex, orthogonality of faces…) strongly impacts SCAO performance. Two different philosophies exist for specifying the precision of the pupil image on the pyramid detector. The first one consists in tightly constraining the optical and mechanical designs in order to ensure a precise positioning of the pupils with respect to the pixels, typically at a $1/10^{th}$ pixel precision. The second consists, on the contrary, in accepting that the absolute pupil positions may be inaccurately distributed (this remaining a static defect), and at the condition of having a reasonable method for detecting the valid pixels. In order to keep a simple and low-cost optical design, we choose to follow the second strategy. We demonstrate in this section with end-to-end simulations, that the pupil position with respect to the pixel grid can be captured by increasing the amount of available pixels and taken it into account in the interaction matrix (i.e. calibration).

The number of available pixels on the P-WFS detector is 240x240. That means that the maximum number per pupil is 120 pixels; the minimum being approx. 80x80 due to the number of actuators used in M4. Figure 12 shows performance in terms of nm RMS and Strehl for the 100x100 case and with one of the pupil shifted along the y-axis as shown on the inset. Two different situations are investigated. One with a calibration matrix (CM) update, corresponding the static aberrations. The other with no CM update, corresponding to a dynamic error in the position of the one of the 4 P-WFS pupils. Figure 13 shows the P-WFS double prism and the 4 P-WFS pupils layout on the detector.

These simulations clearly show that shifts of the P-WFS pupils on the detector can lead to a large decrease in performance. Without calibration update (i.e. dynamic changes), the divergence in SCAO performance start as soon the lateral shift is larger than 0.3 pixels. With calibration (static change) on the other side allows a shift as large as 0.7 pixel without noticeable impact. Actually a shift of 0.7 pixels can be made equivalent to the 0.3 shift; the maximal shift being 0.5. In addition, the AO system performance is only marginally sensitive to the orthogonality of the faces, as long as the interaction matrix takes it into account and with sufficient sampling of P-WFS signals.

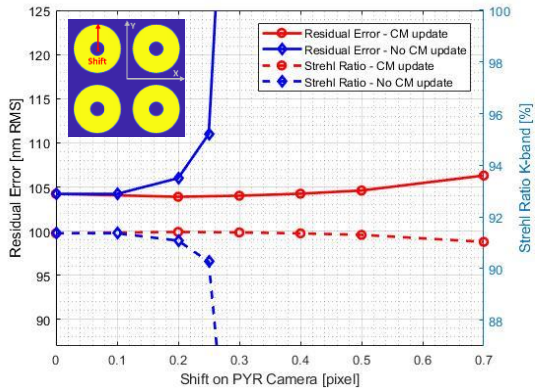

Figure 12: SCAO performance as a function of pixel sift on detector.

Figure 13: Sketch of the P-WFS double prism (right). Position of the 4 P-WFS pupils on the detector (left). The pupil size is 2400 µm, the distance between 2 pupils is 480 µm, and the distance to the edge is 240 µm. The overall dimentions of the OCAM2 detector is 240x240 pixels.

As mentioned previously, the near Nyquist sampling requires 80x80 pixels per P-WFS pupil. Simulations shows that this sampling require exquisite optical quality. Even with new CM calibration, a noticeable drop in performance can be seen for shift larger of equal to 0.3 pixels. Over-sampling the P-WFS signal mitigates greatly these effects while relaxing some of the optical constrains. The only loss in performance will occur for dim natural guide stars, where the light will be spread over more pixels. For these reasons, the SCAO module of HARMONI will use 100x100 per P-WFS pupil.

## 5.2 P-WFS detailed optical design

As mentioned in section 2, the P-WFS module is located between the Focal Plane Relay System & Cryostat between -10 and -20ºC. It requires a large field-of-view for sensing on solar system extended objects, which really imposes the physical size of the optics. The wavelength coverage is set by the dichroic cut-offs between 700-1000 nm. A summary of the top-level requirements is presented Figure 14.

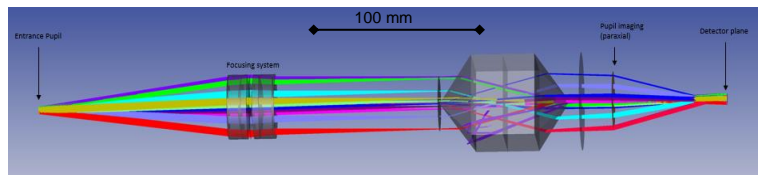

| Top-level Specification | Value |
| --- | --- |
| Field of View | 5 arcsec (goal 6 arcsec) |
| Wavelength | 700 nm to 1000 nm |
| Entrance f/ratio | 30 |
| Environnemental conditions | -10°C (± TBC) |

Figure 14: Top-level requirement for the pyramid optics of HARMONI.

Figure 15: P-WFS simplified optical layout.

Figure 16 shows the P-WFS pupil and manufacturing specifications for the pyramid optics of HARMONI. The current design for the Prism has a chromatism <0.3 µm in the detector plane covering the entire wavelength range (700-1000 nm).

| P-WFS Pupil Specification | Value |
| --- | --- |
| Pupil Diameter | 2400 µm ± 10 µm |
| Pupil diameter dispersion | < 5 µm (at all wavelengths) |
| Lateral Pupil Positions | ±50 µm |

| Manufacturing Specification | Value |
| --- | --- |
| Edge Sharpness | < 4 µm |
| Roof Size | < 10 µm (goal 5 µm) |
| Angular Accuracy | < 5 arcmin |
| Glasses differential CTE | < 1 |

Figure 16: P-WFS pupil (left) and manufacturing (right) specifications for the pyramid optics of HARMONI.

**Designing the Field of View**. The outer beams define the FoV (see the pink and yellow beams in Figure 17). They exit the prisms near the rear edge of the double pyramid by the opposite face. The FoV can be adjusted by changing the thickness both pyramid bases.

**Correction of the chromatism and limit angles**. The choice of glass types is crucial. It influences at the same time chromatism and limit angles. High refractive index is needed to keep the angles small and avoid reflexions. However, we cannot afford too high refractive indices, as we want to minimise the pyramid footprint and keep a good relative dispersion (chromatism).

The thermal environment during the SCAO AIT phase and during operation is also important parameter. The system must keep its performances over a temperature range of -20 to 20°C. The thermal expansion coefficients for both glasses must be very closely matched to allow the two pyramids to be glued together. Moreover, both glasses need to have a low thermal variability of their refractive index. The two glasses that were selected are Ohara S-NSL36 and Ohara S-BAL14 (see Table 4 for further details).

Table 4: Glasses description of achromatic pyramid for HARMONI SCAO

|  | S-NSL36 | S-BAL14 |
|---|---|---|
| Refractive index $n_s$ (@852 nm) | 1.50924 | 1.56040 |
| Abbe Numbe [$v_d$] | 52.43 | 56.36 |
| CTE [µm/m.C°] | 8 | 8 |
| Thermal Conductivity [W/m.K] | 1.09 | 0.967 |
| dn/dT relative [$10^{-6}$/°C] 656 to 1013 nm. -20 to 20 °C | 2 | 1.4 |
| Internal transmittance (@800 nm) | 0.999 | 0.998 |

**Dimensions**. By using this combination of glasses, we obtain 44.6° for the entrance angle and 43.1° for the output angle. The large FoV of 6 arcsec that is required for the SCAO imposes a large footprint for the pyramids: 44x40 mm$^2$ by 82 mm in length. Mechanical and chemical aspects were checked to conform with the specifications required for ELT.

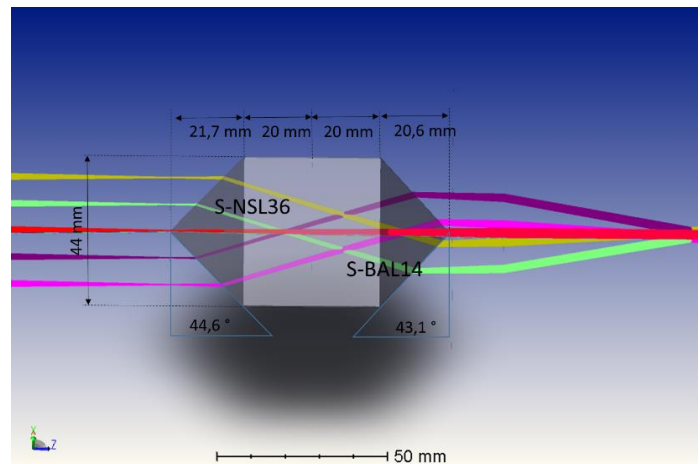

Figure 17: Geometric description of the pyramid of HARMONI SCAO.

The focusing system located before the pyramid (see Figure 15), produces an f/ratio = 30 (4 mm pupil diameter and 120 mm focal distance). It is diffraction limited over a 6 arcsec FoV and a wavelength range between 700 to 1000 nm. Beams are telecentric at the entrance of the pyramid. After the pyramid, the pupil imager must have a 72 mm focal length and keep the image and pupil qualities. Its design is still work in progress (Figure 15).

## 6. CONCLUSIONS

In this paper, we presented simulation results for the single-conjugate AO mode of HARMONI. We have shown that the impact of the final phasing errors of the ELT M1 mirror should not further influence SCAO performance despite their inherent phase discontinuities. We present numerical results for establishing the desired (goal) RTC latency - especially for high contrast – and showed that a latency below two frames at 1 kHz is required. Finally, we focused our study on the P-WFS, investigating the optical gains issues and glass pyramid manufacturing and assembly tolerances. We have shown that oversampling the P-WFS and calibration (i.e. interaction matrix) are able to minimise performance losses due to misalignment and optical errors.

The next step in this investigation is to validate the optical gain model at the scale of the HARMONI SCAO system with an end-to-end simulation. The other aspect is to benchmark new and up-to-date hardware (e.g. the Intel Xeon Platinum 4S) in an attempt to reduce the RTC latency below what has been available to us previously. An alternative solution is to investigate efficient multi-nodes RTC architectures, using the computing hardware necessary for the LTAO mode of HARMONI and left unused during SCAO operations.